# Phonon modes and Raman signatures of MnBi$_{2n}$Te$_{3n+1}$ (*n*=1,2,3,4) magnetic topological heterostructures

Yujin Cho[1*,+], Jin Ho Kang[1,+], Liangbo Liang[2,+], Xiangru Kong[2], Subhajit Ghosh[3], Fariborz Kargar[3], Chaowei Hu[4], Alexander A. Balandin[3], Alexander A. Puretzky[2], Ni Ni[4], and Chee Wei Wong[1,*]

[1]*Fang Lu Mesoscopic Optics and Quantum Electronics Laboratory, University of California, Los Angeles, CA 90095, USA*

[2]*Center for Nanophase Materials Sciences, Oak Ridge National Laboratory, Oak Ridge, TN 37831, USA*

[3]*Phonon Optimized Engineered Materials (POEM) Center, Department of Electrical and Computer Engineering, University of California, Riverside, CA 92521, USA*

[4]*Department of Physics and Astronomy and California Nano Systems Institute, University of California, Los Angeles, CA 90095, USA*

\* yujincho@ucla.edu, cheewei.wong@ucla.edu

[+] Y. Cho, J. H. Kang, and L. Liang contribute equally to this work.

An intrinsic antiferromagnetic topological insulator MnBi$_2$Te$_4$ can be realized by intercalating Mn-Te bilayer chain in a topological insulator, Bi$_2$Te$_3$. MnBi$_2$Te$_4$ provides not only a stable platform to demonstrate exotic physical phenomena, but also easy tunability of the physical properties. For example, inserting more Bi$_2$Te$_3$ layers in between two adjacent MnBi$_2$Te$_4$ weakens the interlayer magnetic interactions between the MnBi$_2$Te$_4$ layers. Here we present the first observations on the inter- and intra-layer phonon modes of MnBi$_{2n}$Te$_{3n+1}$ (*n*=1,2,3,4) using cryogenic low-frequency Raman spectroscopy. We experimentally and theoretically distinguish the Raman vibrational modes using various polarization configurations. The two peaks at 66 cm$^{-1}$ and 112 cm$^{-1}$ show an abnormal perturbation in the Raman linewidths below the magnetic transition temperature due to spin-phonon coupling. In MnBi$_4$Te$_7$, the Bi$_2$Te$_3$ layers induce Davydov splitting of the $A_{1g}$ mode around 137 cm$^{-1}$ at 5 K. Using the linear chain model, we estimate the out-of-plane interlayer force **constant to be $(3.98 \pm 0.14) \times 10^{19}$ N/m³ at 5 K, three times weaker than that of Bi$_2$Te$_3$. Our work discovers the dynamics of phonon modes of the MnBi$_2$Te$_4$ and the effect of the additional Bi$_2$Te$_3$ layers, providing the first-principles guidance to tailor the physical properties of layered heterostructures.**

**Introduction**

The remarkable potential of magnetic topological insulators to realize exotic physical phenomena [1-3] like quantum anomalous Hall effect [4-7] or axion insulators [8-10] have led researchers striving to combine topological properties and magnetism by creating a proximity effect between a topological insulator and magnetic material [11], or by chemical doping [7, 12, 13]. However, synthesizing the material uniformly over a large area with well-controlled doping concentration has proven difficult. The recent discovery of intrinsic antiferromagnetic topological insulator MnBi$_2$Te$_4$ overcomes these difficulties [4, 8, 14, 15]. It can be synthesized into a few millimeters of uniform bulk single crystals and its layered structure exhibits interesting thickness dependence [16]. For example, a structure with an odd number of MnBi$_2$Te$_4$ layers is ferromagnetic, but that with an even number of layers (or bulk) becomes antiferromagnetic due to its head-to-head spin alignment [16-20]. Furthermore, inserting Bi$_2$Te$_3$ layers between two adjacent MnBi$_2$Te$_4$ layers changes the interlayer interactions, leading to altered physical properties such as different magnetic properties [15, 17, 18, 21-24]. The Néel temperature ($T_N$) of bulk MnBi$_2$Te$_4$ is 25 K [19, 20], while for MnBi$_4$Te$_7$ is 13 K [17, 22, 23] and for MnBi$_6$Te$_{10}$ is 11 K [23, 24]. Contrastingly, MnBi$_8$Te$_{13}$ becomes ferromagnetic below 10.5 K [21]. Their bulk properties have been studied by various methods, including transport measurements [19, 21-24], angle-resolved photoemission spectroscopy [17, 20-22, 25], *x*-ray scattering [19, 23], and magnetic susceptibility measurements [17, 19, 22, 23].

Here we present the study of phonon vibrational modes in bulk MnBi$_{2n}$Te$_{3n+1}$ (*n*=1,2,3,4) using linear- and circularly-polarized cryogenic low-frequency optical Raman spectroscopy, supported by first-principles density functional theory (DFT) calculations. Raman spectroscopy is non-destructive and highly sensitive to small changes in the lattice dynamics and magnetic ordering down to the two-dimensional limit [19, 26-30]. This work provides a thorough examination and physical understanding of intralayer phonon dynamics in MnBi$_{2n}$T$_{3n+1}$ (*n*=1,2,3,4) and interlayer interactions, in support of magnetic topological insulator studies at the few-layers limit.

**Results and Discussion**

Figure 1a, left panel, depicts the side view of the MnBi$_2$Te$_4$ crystal structure. One unit layer consists of seven atomic layers (a septuple layer) and is a compound of a Bi$_2$Te$_3$ and a Mn-Te bilayer chain, of which the latter is shown in the dashed box. The top view of MnBi$_2$Te$_4$ has three-fold symmetry as illustrated in the middle panel of Figure 1a [15]. The layered structures of MBT-families are shown in Figure 1b, which are equivalent to heterostructures with a MnBi$_2$Te$_4$ and a differing number of Bi$_2$Te$_3$. For example, MnBi$_4$Te$_7$, MnBi$_6$Te$_{10}$, and MnBi$_8$Te$_{13}$ have one, two, and three Bi$_2$Te$_3$ layers between two MnBi$_2$Te$_4$ layers, respectively. Figure 1c shows an example Raman spectrum of a bulk MnBi$_2$Te$_4$ at 7 K (top) and the calculated peak positions from the DFT (bottom). According to group theory, the crystal structure of bulk MnBi$_2$Te$_4$ belongs to the space group $R\bar{3}m$ (No. 166, point group D$_{3d}$) [31-33], and it has doubly degenerate $E_g$ and non-degenerate $A_{1g}$ symmetry Raman modes. To experimentally verify the *E*- and *A*-modes of the Raman peaks, we use circular polarization, which can be found in Figure S1 in Supplementary Information (SI). From the Raman tensor analysis provided in SI [34], *E*-modes only appear in $\bar{z}(\sigma^- - \sigma^+)z$ configuration, while *A*-modes are only present in $\bar{z}(\sigma^- - \sigma^-)z$ configuration. From our measurements, we assign Peaks II and III to *E*-modes (red) and Peaks I, IV, and V to *A*-modes (blue), consistent with DFT calculations where red (blue)-lines are *E* (*A*)-modes. The gray shaded area indicates the cutoff frequency in the polarized cryogenic Raman setup. The lowest frequency of the *E*-modes of 30.3 cm$^{-1}$ from theory was measured with Horiba T64000 Raman spectrometer in ambient, shown in Figure S2 in SI.

In Figure 1d, we present the incident linear polarization dependence of MnBi$_2$Te$_4$ at 7 K, maintaining the relative angle of the analyzer in the parallel polarization configuration. The polarization dependence remains isotropic, consistent with our theoretical analysis for the $R\bar{3}m$ space group without considering the spin ordering (more details in SI). This means no observable in-plane magnetic anisotropy due to spin-phonon coupling in the antiferromagnetic state. Vibrational modes and the temperature dependence of MnBi$_2$Te$_4$ are discussed in depth later.

Figure 2 presents the Raman spectra of MnBi$_{2n}$Te$_{3n+1}$ (*n*=1,2,3,4) at room temperature. Each sample has (*n*-1) number of Bi$_2$Te$_3$ layers between two MnBi$_2$Te$_4$ layers. In all samples, we observe four distinctive peaks around 42 (peak I), 60 (II), 100 (III), and 135 cm$^{-1}$ (V), and a shoulder peak around 110 cm$^{-1}$ (IV). The peak marked with * comes from the laser. As *n* increases, the peak positions and the widths converge to those of Bi$_2$Te$_3$; light horizontal lines in Figures 2c to 2d). Compared to the Raman spectrum of Bi$_2$Te$_3$, Peaks I, II, and IV uniquely exist in MnBi$_2$Te$_4$. Our DFT calculations show that the Mn atom of the Mn-Te bilayer chain remains static in all vibrational modes, while the Te atom oscillates in- or out-of-plane in Peaks I, II, and IV; see the inset in Figure 2b for peak III as an example and all the vibrational modes in Figure S3a in SI. When $n \geq 2$, the additional Bi$_2$Te$_3$ layer either introduces a new peak or shifts the existing peaks (Figure S3b in SI). For example, the main contribution to the $E_g$ mode of Peak III in MnBi$_2$Te$_4$ comes from the vibrations of the top and the bottom Bi-Te atomic layers, which is like that of Bi$_2$Te$_3$. Therefore, the additional Bi$_2$Te$_3$ layer simply shifts the peak position, instead of creating a new mode.

In Peak II, the Bi$_2$Te$_3$ layer generates a new peak when it oscillates out-of-plane ($A_{1g}$) while the MnBi$_2$Te$_4$ layer remains almost static (calculated frequency of 58.8 cm$^{-1}$ in Figure S3b, SI). This peak position overlaps with the $E_g$ vibration of MnBi$_2$Te$_4$ where the Bi$_2$Te$_3$ layer stays static (calculated frequency of 68.9 cm$^{-1}$ in Figure S3b, SI). Therefore, Peak II with $E_g$ symmetry in the pure MnBi$_2$Te$_4$ bulk develops into two mixed peaks in the MnBi$_2$Te$_4$/Bi$_2$Te$_3$ heterostructure, i.e., one peak still with $E_g$ symmetry with the MnBi$_2$Te$_4$ layer vibrating, and the other one with $A_{1g}$ symmetry with the Bi$_2$Te$_3$ layer vibrating.

We use a circular polarization configuration to experimentally confirm the symmetries. In Figure 2b, the spectrum, measured in $\sigma^- - \sigma^+$ circular polarization, shows only the *E*-modes. We observe that Peak II in MnBi$_2$Te$_4$ redshifts with increasing *n*. The decreased intensity is due to the reduced number of vibration entities in the unit area, owing to the static Bi$_2$Te$_3$ layers in the vibration patterns of Peak II. The dominating intensity around 60 cm$^{-1}$ above $n = 2$ in Figure 2a is from the $A_{1g}$ mode of Bi$_2$Te$_3$, which is absent in MnBi$_2$Te$_4$, as expected from our theory. Polarization measurements in the parallel, cross, and $\sigma^- - \sigma^-$ polarization configurations are shown in Figure S4 in SI, to further differentiate these two mixed peaks. The evolution of the peak positions and the widths with *n* is presented in Figures 2c and 2d respectively. The exact Raman peak positions from the experiment and the theory are given in the Table S1, SI. To observe the evolution of the peak II (orange dashed line), we fit the $E_g$ mode with Lorentzian in Figure 2b, where the $A_{1g}$ mode from the Bi$_2$Te$_3$ layer is suppressed. The orange solid line represents the $A_{1g}$ mode from the Bi$_2$Te$_3$. As expected, the peaks get closer to that of Bi$_2$Te$_3$ with increasing *n*. In this measurement, we observe a Davydov splitting of Peak V on MnBi$_4$Te$_7$ (blue dashed line, $n = 2$), which will be discussed further in Figure 4.

**Temperature-dependent Raman spectroscopy.** To understand the effect of magnetic ordering on Raman spectrum of MnBi$_2$Te$_4$, we examine the temperature-dependent spectra from 294 K to 5 K. Figure 3a shows no new peaks below the Néel temperature, indicating that the magnetic ordering does not significantly change the vibrational modes. Figure 3b displays selected Raman peak shifts of II($E_g^2$), IV($A_{1g}^2$) and V($A_{1g}^3$), relative to those at room temperature. The rest can be found in SI, Figure S5. The Raman shifts follow the general temperature-dependent anharmonicity model based on phonon decays into different number of phonons [35], i.e., monotonic blueshift down to ≈ 40 K, followed by almost constant peak positions at lower temperatures. This model describes our experimental data well, represented by solid lines. In the same model, we would expect the widths of the Raman peaks to get narrower at lower temperatures, as observed in Peaks I, III, and V (e.g., Figure 3d for Peak V and Figure S5 in SI for Peaks I and III) [35]. Interestingly, we find that Peaks II and IV behave differently. Peak IV widens by ≈ 20% at low temperatures, as in Figure 3d, green scatter points. Peak II becomes broader below 20 K, while the width barely changes above 20 K, as in Figure 3e. This behavior is discordant with the anharmonicity model. For Peaks I, III, and V, the main contribution comes from the vibrations of the top and the bottom Bi-Te chains, not directly bonded to Mn atoms [Figure 3c, right panel, and Figure S3a]. Conversely, the main contribution to Peaks II and IV comes from the vibrations of the Te atoms that are directly neighboring with Mn atoms [left two panels in Figure 3c]. Therefore, below the $T_N$, the spin-phonon coupling would be more pronounced in Peaks II and IV, matching our measurements. We further confirmed the AFM ordering with DC magnetic susceptibility measurement, as shown in Figure 3e, right axis. The magnetic susceptibility peaks at 23.3 K, which is a signature of AFM. We do not expect the order parameter-like temperature dependence here, because the net magnetization is close to zero in the ordered AFM phase. Hence, we attribute the observed broadening of the peaks, and therefore decreased phonon lifetime, to the spin-phonon coupling [26, 36, 37]. The temperature from which the Raman peak broadens is slightly lower than the $T_N$. This is possibly because the change in the width is too small near the $T_N$ and falls within the experimental uncertainties. In MnBi$_4$Te$_7$, we also observed a broader width of Peak IV (Figure S6 in SI), but could not isolate the $E_g$ mode from the MnBi$_2$Te$_4$ vibration due to dominating $A_{1g}$ mode from the Bi$_2$Te$_3$ layer near the same frequency.

**Davydov splitting in MnBi$_4$Te$_7$.** When a Bi$_2$Te$_3$ layer forms a heterostructure with MnBi$_2$Te$_4$, it not only introduces an additional mode from Bi$_2$Te$_3$, e.g. $A_{1g}$ mode at 60 cm$^{-1}$, but also perturbs an intralayer vibration mode in bulk MnBi$_2$Te$_4$ and splits a peak, known as Davydov splitting [38], as presented in Figure 4. This Davydov splitting has been observed in other layered materials, such as MoSe$_2$ [39] and MoTe$_2$ [40]. Figure 4a shows the zoomed-in temperature-dependent spectra near Peak V, 135 cm$^{-1}$. The full spectrum can be found in SI, Figure S6. At room temperature, the broad linewidths make the splitting less distinct. The narrower linewidths at cryogenic temperatures reveal Davydov splitting and the shoulder clearly. Figure 4b summarizes the central frequencies of the split from the two Lorentzian fits (purple and green in Figure 4a) at 136.6 cm$^{-1}$ and 140.2 cm$^{-1}$ at 5 K. The $\approx$ 4 cm$^{-1}$ peak separation slightly widens at room temperature (132.0 cm$^{-1}$ and 136.3 cm$^{-1}$), as can be seen on the right axis (grey scatter points) in panel (b).

From our calculations, the $A_{1g}$ mode of MnBi$_2$Te$_4$ at 144.1 cm$^{-1}$ (138.6 cm$^{-1}$ peak in the experiment) has its Raman-inactive Davydov pair at 146.0 cm$^{-1}$. In MnBi$_4$Te$_7$, the additional Bi$_2$Te$_3$ has a similar, but not identical, vibration pattern to the $A_{2u}$ mode in MnBi$_2$Te$_4$; the intercalated Mn-Te chain alters the Bi$_2$Te$_3$ crystallographic structure. The symmetry breaking by the additional Bi$_2$Te$_3$ layer renders this mode Raman active in MnBi$_4$Te$_7$ and induces the $A_{1g}$ peak at 143.4 cm$^{-1}$ (see Figure 4c). The $A_{1g}$ mode of MnBi$_2$Te$_4$ at 144.1 cm$^{-1}$ then downshifts to 141.3 cm$^{-1}$ in MnBi$_4$Te$_7$. Therefore, a single $A_{1g}$ peak in MnBi$_2$Te$_4$ appears as two $A_{1g}$ peaks in the Raman spectrum of MnBi$_4$Te$_7$, corroborating our experimental observations. In $A_1$-like vibrational mode, the vibrational frequencies of these two peaks are related as $\omega_1^2 = \omega_0^2 + \omega_{LB}^2$ where $\omega_0$ is the original $A_{1g}$ peak (136.6 cm$^{-1}$; experiment at 5 K) and $\omega_1$ is its Davydov pair (140.2 cm$^{-1}$; experiment at 5 K) [41, 42]. $\omega_{LB}$ is the frequency of layer breathing (LB) mode. From this expression, we experimentally find that $\omega_{LB}$ is 31.7±0.6 cm$^{-1}$ at 5 K and 33.7±2.5 cm$^{-1}$ at 297 K. Theoretically, we estimate the $\omega_{LB}$ ($A_{2u}$ mode; Raman inactive) to be 22.48 cm$^{-1}$ which deviates from our measurements due to the overestimated peaks from the DFT calculations. Using the linear chain model [40, 41, 43], we can estimate the out-of-plane interlayer force constant ($K^\perp$) using $\omega_{LB} = (1/\pi c)\sqrt{K^\perp/\mu}$; $c$ is the speed of light and $\mu$ is the effective mass per unit area. The estimated interlayer $K^\perp$ in MnBi$_4$Te$_7$ thus ranges from $(3.98 \pm 0.14) \times 10^{19}$ N/m$^3$ to $(4.48 \pm 0.66) \times 10^{19}$ N/m$^3$ between 5 K and room temperature, which is about a third of the value in Bi$_2$Te$_3$ [41]. In MnBi$_6$Te$_{10}$ and MnBi$_8$Te$_{13}$, it is difficult to observe such a splitting at this excitation wavelength because dominating Bi$_2$Te$_3$ vibrational modes obscure the Davydov splitting.

**Conclusions.** We examined the phonon dynamics of the layered MnBi$_{2n}$Te$_{3n+1}$ magnetic topological insulator using polarized low-frequency Raman spectroscopy at cryogenic temperatures supported by first-principle DFT calculations of the phonon modes. Comparison of MnBi$_2$Te$_4$ and MnBi$_4$Te$_7$ provides us the best contrast in the Raman phonon spectra due to a newly introduced Bi$_2$Te$_3$ layer. From MnBi$_4$Te$_7$ to MnBi$_8$Te$_{13}$, the spectra change more gradually. In MnBi$_2$Te$_4$, the anomalous broadening of the 66 cm$^{-1}$ peak is observed below the Néel temperature, the origin of which is explained by the intralayer spin-phonon coupling. When a Bi$_2$Te$_3$ layer is inserted between two MnBi$_2$Te$_4$ layers, an out-of-plane vibrational mode from Bi$_2$Te$_3$ dominates Peak II. We also observed and quantified the Davydov splitting of $A_{1g}$ mode at 136.6 cm$^{-1}$, arising from interlayer interactions between Bi$_2$Te$_3$ and MnBi$_2$Te$_4$. The estimated out-of-plane interlayer force constant is between $(3.98 \pm 0.14)$ to $(4.48 \pm 0.66) \times 10^{19}$ N/m$^3$, at 5 K and 294 K, respectively, which is weaker than that of Bi$_2$Te$_3$ by a factor of 3. Our studies advance the understanding of the phonon modes and layer interactions in MnBi$_{2n}$Te$_{3n+1}$ magnetic topological insulators, including circularly polarized excitations, anharmonic shifts and broadening, spin-phonon coupling, and Davydov splitting. It unveils the role of the intercalated Mn-Te bilayers in $Bi_2Te_3$ and supports the control of the physical processes in heterogeneous magnetic topological insulators.

**Methods**

**Sample preparation:** We synthesize $MnBi_{2n}Te_{3n+1}$ ($n$=1,2,3,4) single crystals with $Bi_2Te_3$ flux [21, 22, 25, 31]. Mn, Bi, and Te elements, at a molar ratio of 15:170:270, are loaded in a crucible and sealed in a quartz tube under one-third atmospheric pressure of Ar. The ampule is heated to 900°C for 5 hours, followed by air-quenching. It is then transferred to another furnace to cool it slowly from 595°C to the decanting temperature, which varies for different $n$ [21]. After resting the samples for one day, we centrifuge the ampule to separate the $MnBi_{2n}Te_{3n+1}$ crystals from the flux. To achieve clean and flat surfaces for the optical measurements, we mechanically exfoliate the bulk single crystals on Si substrates. The average thickness of the flakes is thicker than 100 nm. All crystallographic structure images were produced by VESTA [44].

**Raman spectroscopy:** We use a 532.2 nm laser, with an 80× objective lens at a power below 1 mW. Two Ondax notch filters suppress the laser Rayleigh scattering and reveal low-frequency Raman peaks down to 30 cm$^{-1}$. The Raman spectrum is acquired with a Horiba 1000M spectrometer with 1200 grooves/mm grating and the spectrometer was calibrated with a Mercury lamp and Si peak at 520 cm$^{-1}$ before the measurements. To ensure the repeatability of the Raman signal, we probe several flakes, whose spectrum is consistent within the experimental uncertainties. No laser-induced effect is observed over a few hours.

**Density Functional Theory calculations:** We perform spin-polarized density functional theory (DFT) calculations using software VASP [45]. Projector-augmented-wave (PAW) pseudopotentials are used to capture the electron-ion interactions and generalized gradient approximation (GGA) with the Perdew-Burke-Ernzerhof (PBE) functional [46] to understand the exchange-correlation interactions. The GGA+U method takes into account the localized $d$-orbitals of Mn atoms, with the U parameter to be 4 [15] and the DFT-D3 method [47] describes the van der Waals interactions. We relax the lattice constants and the atomic positions of bulk $MnBi_2Te_4$ and $MnBi_2Te_4/Bi_2Te_3$ heterostructure ($MnBi_4Te_7$), until the residual forces fall below 0.001 eV/Å. The cutoff energy is set at 350 eV and the *K*-point mesh was 12×12×3. On the optimized unit cell, we then perform phonon calculations using a finite difference scheme implemented in the Phonopy software [48]. We use VASP to compute Hellmann-Feynman forces in the 3×3×1 supercell for both positive and negative atomic displacements ($\delta$ = 0.03 Å), and Phonopy to construct the dynamic matrix whose diagonalization provides phonon frequencies and phonon eigenvectors such as the phonon vibration mode patterns.

**Data Availability:** The datasets generated during and/or analyzed during the current study are available from the corresponding authors on reasonable request.

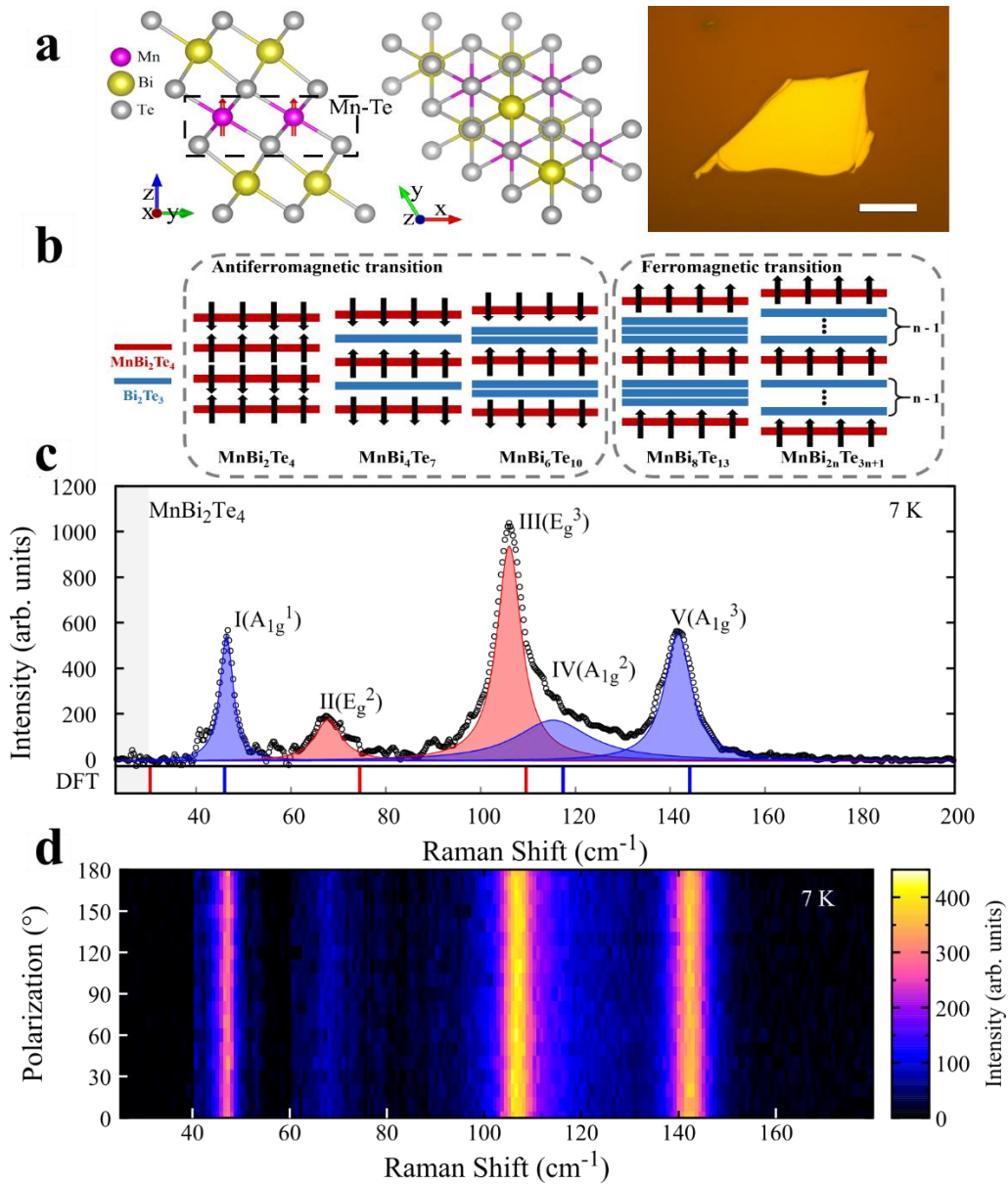

**Figure 1 | Polarized optical Raman spectroscopy of magnetic topological insulators, distinguishing *E*-mode and *A*-mode phonons. a,** Crystal structure of $MnBi_2Te_4$ (left), top view of $MnBi_2Te_4$ (middle), and exemplary optical image of samples (right). Scale bar: 20 µm. **b,** The figure shows layered structure of MBT-families. **c,** shows Raman spectrum (top) of $MnBi_2Te_4$ in *E*-mode (red) and *A*-mode (blue), measured at 7 K. DFT calculation (bottom) shows corresponding Raman peak positions in each polarization configuration. The grey shaded area below 30 cm$^{-1}$ indicates the cut-off filter frequency. **d,** Linear parallel polarization dependence of Raman signal at 7 K.

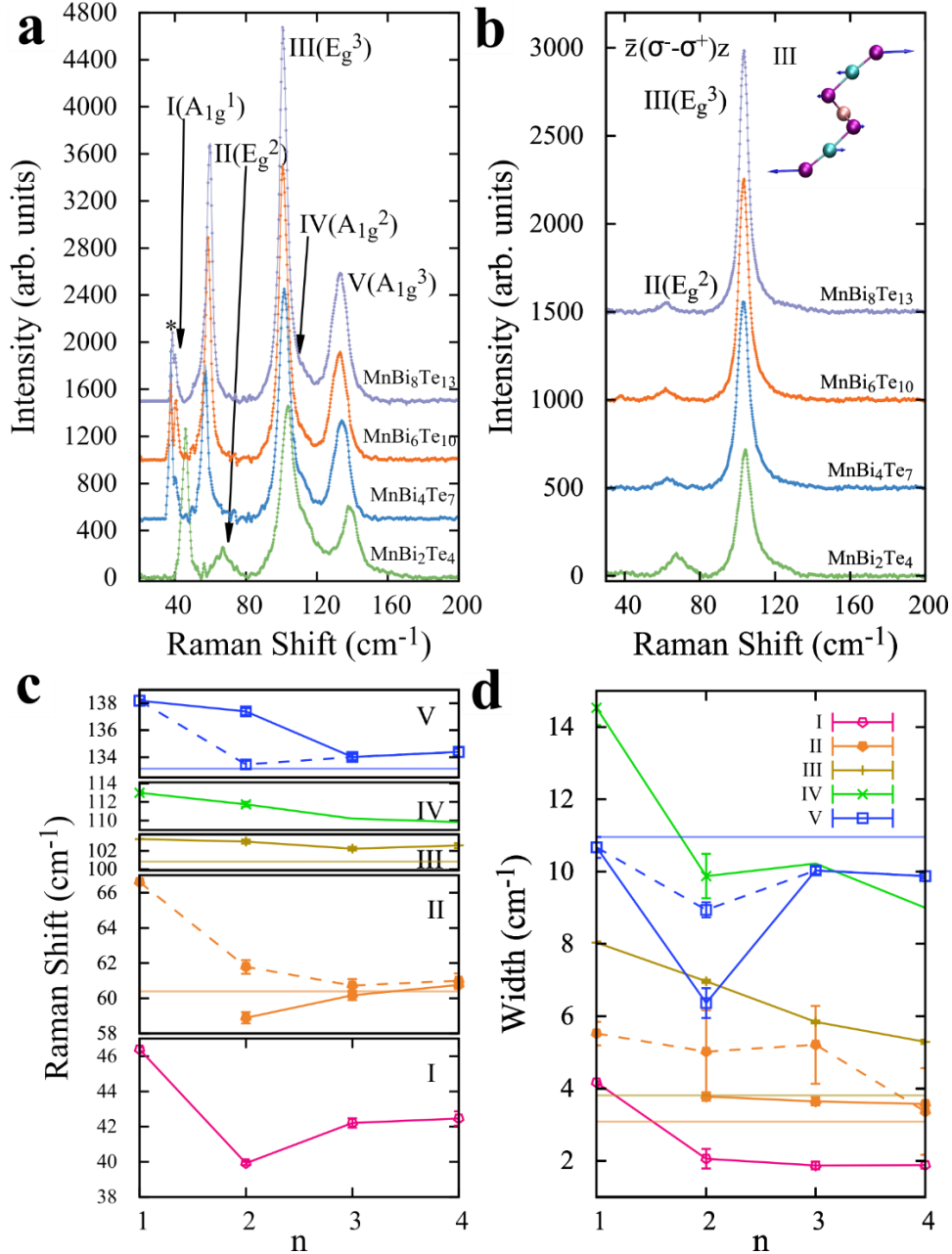

**Figure 2 | Effects of Bi$_2$Te$_3$ intercalation on Raman spectra of MnBi$_{2n}$Te$_{3n+1}$ topological heterostructures. a** and **b**, Raman spectra measured at room temperature in linear polarization configuration without analyzer and with circularly polarized light $[\bar{z}(\sigma^- - \sigma^+)z]$, respectively. Inset shows example of in-plane vibration of structure corresponding to peak III. **c** and **d**, Evolution of Raman peaks positions and widths, respectively. Light horizontal lines indicate peak positions and widths of bulk Bi$_2$Te$_3$ measured at the same temperature.

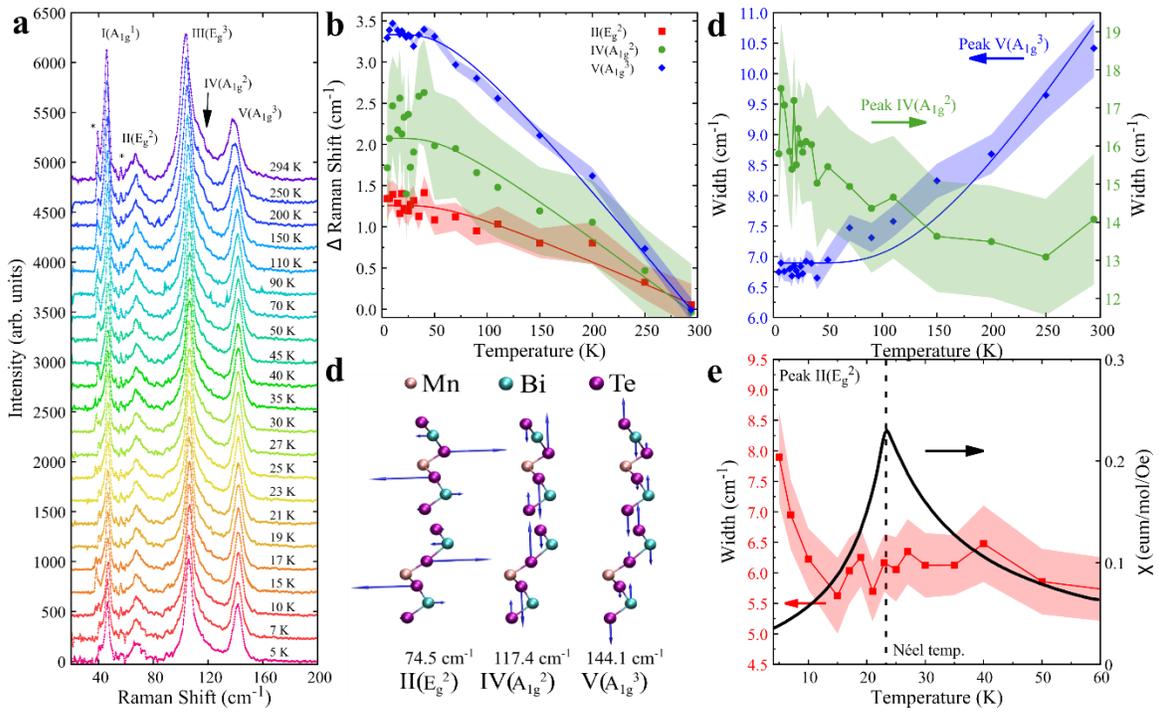

**Figure 3 | Temperature dependences and intralayer spin-phonon coupling of antiferromagnetic MnBi$_2$Te$_4$ topological insulator. a,** Raman spectra measured in temperatures ranging from 5 to 294K. Peaks marked with * are from laser. **b,** Change in Raman peak positions of II(E$_g^2$), IV(A$_{1g}^2$) and V(A$_{1g}^3$), relative to those measured at room temperature. Solid lines represent fitting based on anharmonicity model. **c,** Vibrational patterns of II, IV, and V modes. **d,** Widths of peak V(A$_{1g}^3$) and peak IV(A$_{1g}^2$), obtained from Lorentzian fits. Blue solid line is fit with anharmonicity model. **e,** Temperature dependent Raman linewidth of Peak II (red scatter points, left axis) and magnetic susceptibility of bulk MnBi$_2$Te$_4$ (black solid line, right axis). The shaded area in panel (**b**, **c**, and **e**) indicates the experimental errors.

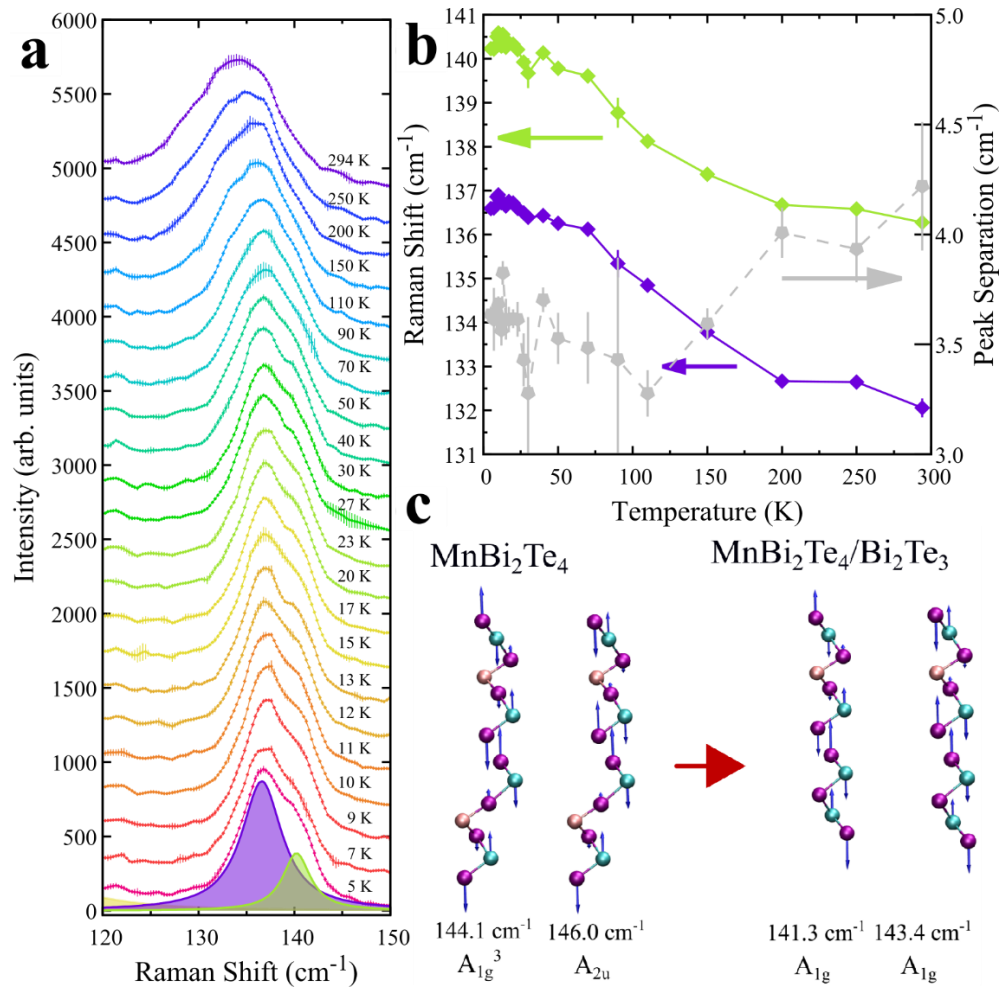

**Figure 4 | Intralayer Davydov splitting from interlayer interactions in MnBi$_4$Te$_7$ heterostructures. a,** Temperature dependences of MnBi$_4$Te$_7$, focusing on Peak V centered at 135 cm$^{-1}$. Two Lorentzian fits for Peak V are shown for spectrum at 5 K. **b,** Evolution of the splitting Raman peaks (left axis; green and purple) and the peak separation between the Davydov pair (right axis; grey). **c,** Change in vibrational modes between MnBi$_2$Te$_4$ and MnBi$_4$Te$_7$ (MnBi$_2$Te$_4$/Bi$_2$Te$_3$ heterostructure) based on DFT calculations.

# Supplementary Information

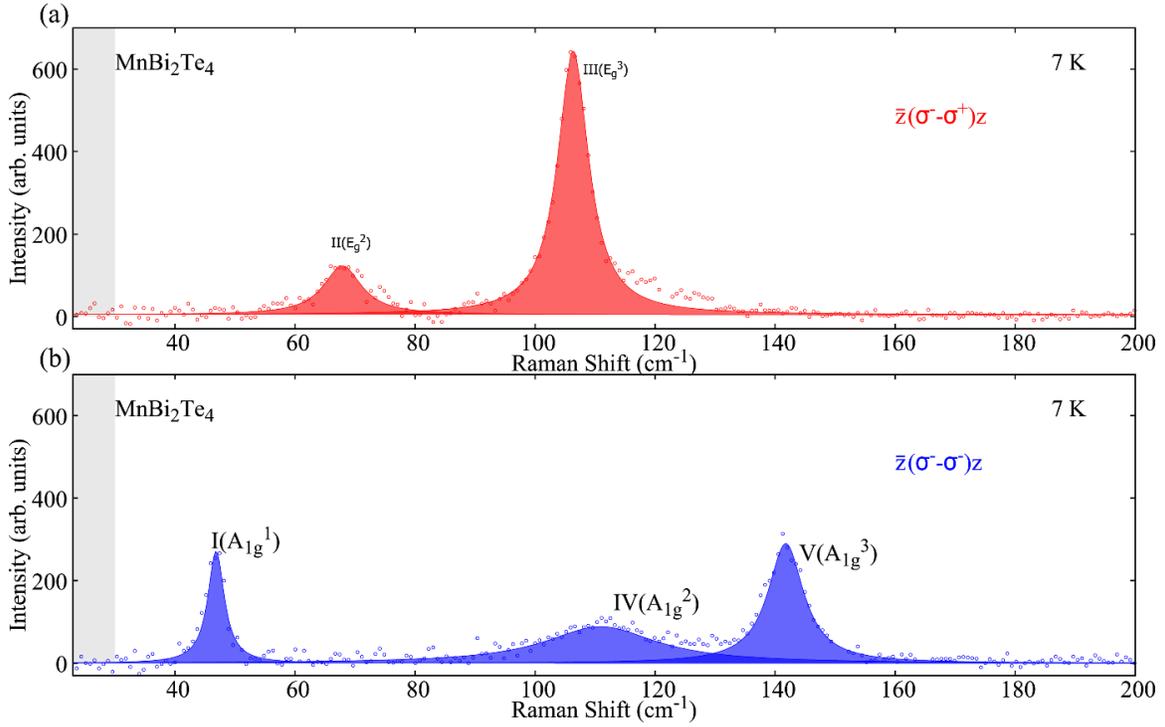

**Figure S1.** Circularly polarized optical Raman spectroscopy of magnetic topological insulators, distinguishing $E$-mode $[\bar{z}(\sigma^- - \sigma^+)z]$ (a) and $A$-mode $[\bar{z}(\sigma^- - \sigma^-)z]$ (b) phonons at 7 K. Shaded area indicates the cut-off filter frequency.

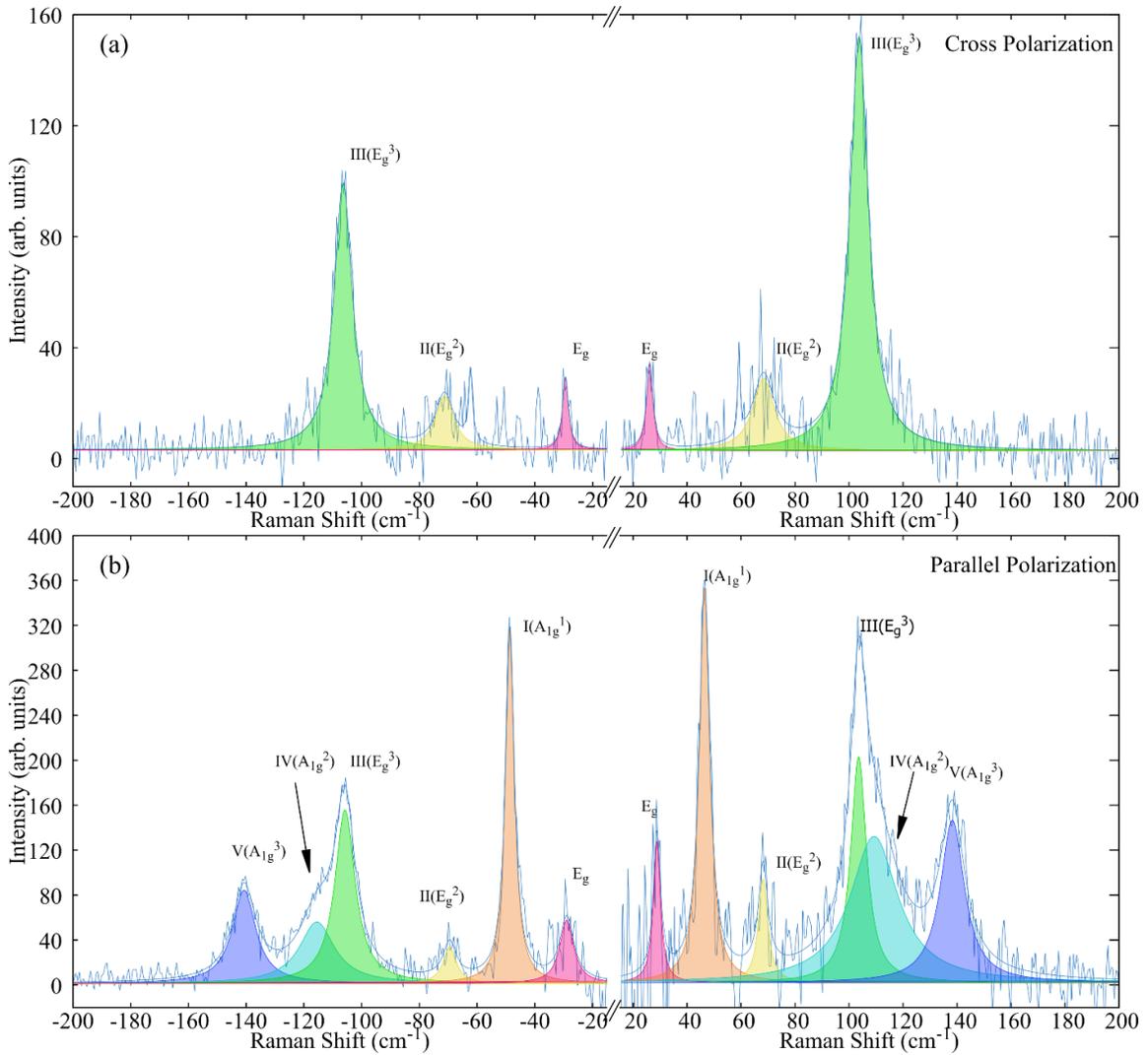

**Figure S2.** Stokes and Anti-Stokes Raman spectrum of MnBi$_2$Te$_4$ in linear polarization configuration in ambient. (a) the Raman spectrum in cross linear polarization selects only *E*-modes. (b) both *A*- and *E*-modes appear in parallel linear polarization. We use Horiba T64000 with 532.2 nm laser in ambient condition at Oak Ridge National Laboratory for Stokes and anti-Stokes.

- **Raman group theory analysis for MnBi$_2$Te$_4$ and MnBi$_4$Te$_7$**

According to group theory analysis, bulk MnBi$_2$Te$_4$ belongs to the space group $R\bar{3}m$ (No. 166) with the point group D$_{3d}$, and it has doubly degenerate $E_g$ symmetry Raman modes ($E_g - x$ and $E_g - y$) and non-degenerate $A_{1g}$ symmetry Raman modes. Their Raman tensors $\tilde{R}$ are

$$\tilde{R}(E_g - x) = \begin{pmatrix} c & \cdot & \cdot \\ \cdot & -c & d \\ \cdot & d & \cdot \end{pmatrix}, \tilde{R}(E_g - y) = \begin{pmatrix} \cdot & c & d \\ c & \cdot & \cdot \\ d & \cdot & \cdot \end{pmatrix};$$

$$\tilde{R}(A_{1g}) = \begin{pmatrix} a & \cdot & \cdot \\ \cdot & a & \cdot \\ \cdot & \cdot & b \end{pmatrix}. \tag{S1}$$

- **Linear Polarization**

In the experimental back-scattering laser geometry (light Z in and Z out) with linear polarization, the electric polarization vectors of the incident and scattered light $e_i$ and $e_s$ are in-plane (i.e., the X-Y plane), and they are given by $e_i = (cos\theta, sin\theta, 0)$ and $e_s = (cos\gamma, sin\gamma, 0)$. With Raman intensity $I \propto |e_i \cdot \tilde{R} \cdot e_s^T|^2$, we have

$$I \propto \left| (cos\theta, sin\theta, 0) \quad \tilde{R} \quad \begin{pmatrix} cos\gamma \\ sin\gamma \\ 0 \end{pmatrix} \right|^2. \tag{S2}$$

Substituting the Raman tensors $\tilde{R}$ from Eq. S1 into Eq. S2, we can obtain

$$I(E_g - x) \propto |c|^2 cos^2(\theta + \gamma), I(E_g - y) \propto |c|^2 sin^2(\theta + \gamma);$$

$$I(A_{1g}) \propto |a|^2 cos^2(\theta - \gamma).$$

Consequently, the intensity of a doubly degenerate $E_g$ Raman peak is

$$I(E_g) = I(E_g - x) + I(E_g - y) = |c|^2 cos^2(\theta + \gamma) + |c|^2 sin^2(\theta + \gamma) = |c|^2,$$

which indicates that the polarization profile of an $E_g$ peak under any linear polarization configuration should be independent of the polarization angle and thus isotropic, consistent with our experimental observation.

For an $A_{1g}$ Raman peak, on the other hand, its intensity is generally dependent on the polarization angle; however, under the experimental parallel polarization configuration, $\gamma = \theta$, we have $I(A_{1g}) \propto |a|^2$, which shows that the polarization profile of an $A_{1g}$ peak under the parallel polarization configuration is also isotropic, consistent with the experimental data.

- **Circular Polarization**

When we use incident circular polarization and measure circularly polarized Raman signal, the Raman intensity $I$ is proportional to $|\sigma_s^\dagger \cdot \tilde{R} \cdot \sigma_i^T|^2$ where $\sigma_s$ and $\sigma_i$ are the polarization vectors of the scattered and incident circularly polarized light, respectively. The left and right circular-polarization vectors can be written as $\sigma_\pm = \frac{1}{\sqrt{2}}(1, \mp i, 0)$ where $\sigma_+^\dagger = \sigma_-$ and $\sigma_-^\dagger = \sigma_+$. Under the same circular-polarization incident and scattered light, the Raman intensity is

$$I \propto \left| \frac{1}{2}(1, i, 0) \quad \tilde{R} \quad \begin{pmatrix} 1 \\ -i \\ 0 \end{pmatrix} \right|^2 \quad \text{when } \sigma_s = \sigma_i = \sigma_+$$

and $$I \propto \left| \frac{1}{2}(1, -i, 0) \quad \tilde{R} \quad \begin{pmatrix} 1 \\ i \\ 0 \end{pmatrix} \right|^2 \quad \text{when } \sigma_s = \sigma_i = \sigma_- .$$

According to the Raman tensors $\tilde{R}$ from Eq. S1, in both cases, the Raman intensity for the $A_{1g}$ mode becomes $I(A_{1g}) \propto |a|^2$, while the intensity for the $E_g$ mode becomes zero. On the contrary, under the opposite circular-polarization incident and the scattered light, the Raman intensity is

$$I \propto \left| \frac{1}{2}(1, i, 0) \quad \tilde{R} \quad \begin{pmatrix} 1 \\ i \\ 0 \end{pmatrix} \right|^2 \quad \text{when } \sigma_s = -\sigma_i = \sigma_+$$

and $$I \propto \left| \frac{1}{2}(1, -i, 0) \quad \tilde{R} \quad \begin{pmatrix} 1 \\ -i \\ 0 \end{pmatrix} \right|^2 \quad \text{when } \sigma_s = -\sigma_i = \sigma_- .$$

In both cases, only the Raman intensity of the $E_g$ mode becomes non-zero.

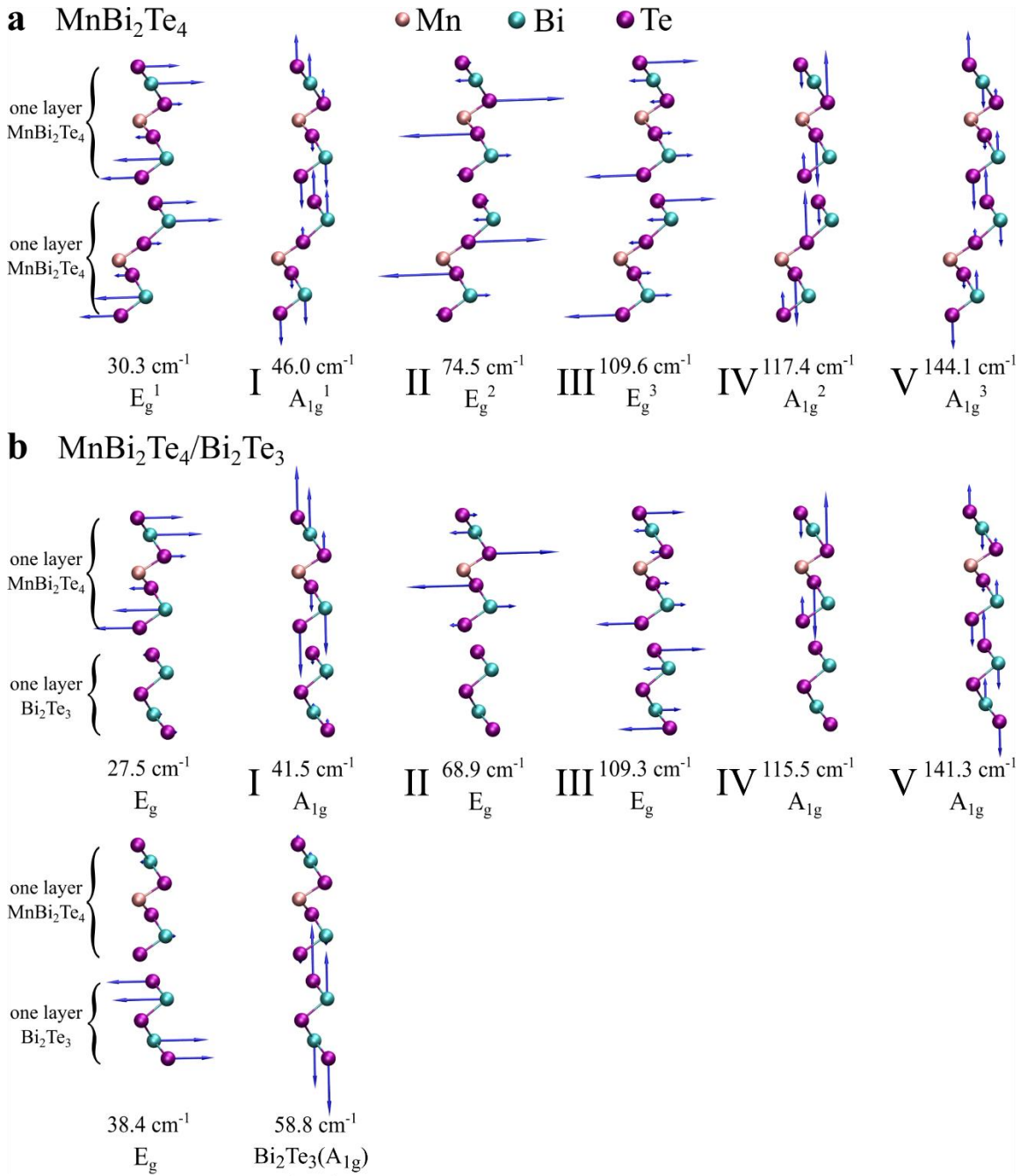

**Figure S3.** Calculated phonon vibration modes of MnBi$_2$Te$_4$ (panel a) and MnBi$_2$Te$_4$/Bi$_2$Te$_3$ heterostructures (panel b). The blue arrows indicate the strength of the vibration of each atom. The calculated phonon frequency and symmetry assignment are shown below each vibration pattern. We note that DFT calculations can overestimate the frequencies by 5 to 10 cm$^{-1}$, compared to experimental values. We note that DFT calculations can overestimate the peak positions by 5 to 10 cm$^{-1}$.

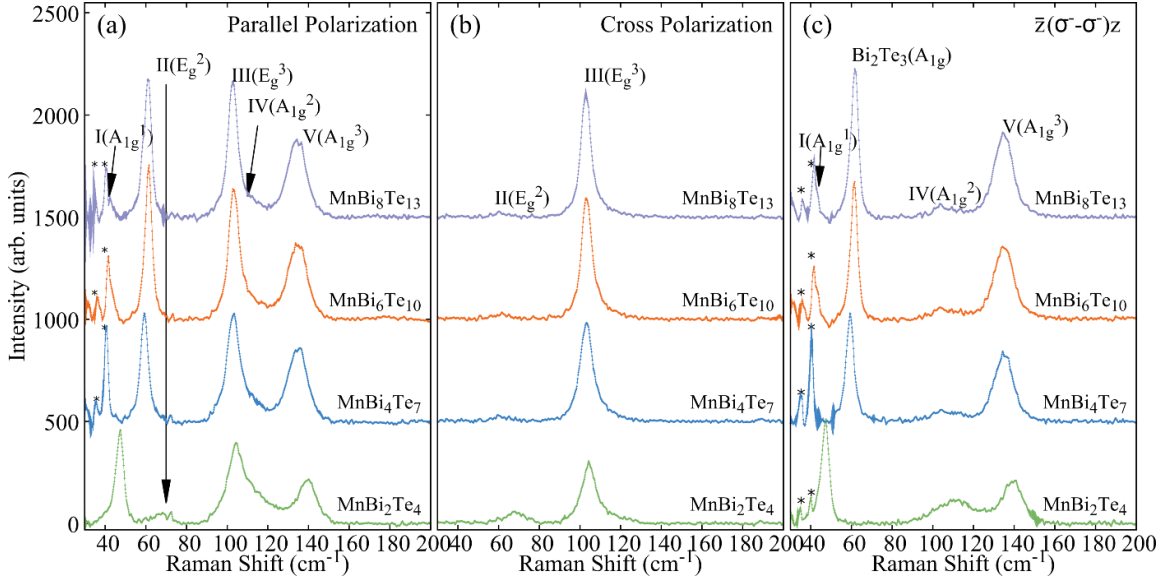

**Figure S4.** Raman spectra of MnBi$_{2n}$Te$_{3n+1}$ ($n$=1,2,3,4) at room temperature, measured in a parallel-polarization (panel a), a cross-polarization (panel b), and $\sigma^- - \sigma^-$ polarization (panel c) configurations. The peak marked with * is from the laser line. Note that Peak II with $E_g$ symmetry in the pure MnBi$_2$Te$_4$ bulk develops into two mixed peaks in the MnBi$_2$Te$_4$/Bi$_2$Te$_3$ heterostructure: one still with $E_g$ symmetry with the MnBi$_2$Te$_4$ layer vibrating and thus still denoted as Peak II ($E_g$) in the heterostructure, while the other one with $A_{1g}$ symmetry with the Bi$_2$Te$_3$ layer vibrating and therefore denoted as Bi$_2$Te$_3$($A_{1g}$) in the heterostructure (see Figure S3 for the vibrational patterns). These two peaks are located around 60 cm$^{-1}$, and we can use linear and circular polarization to differentiate them. A peak with $E$ symmetry can appear in the parallel-polarization, cross-polarization, and $\sigma^- - \sigma^+$ polarization; on the other hand, a peak with $A$ symmetry appears in the parallel-polarization and $\sigma^- - \sigma^-$ polarization.

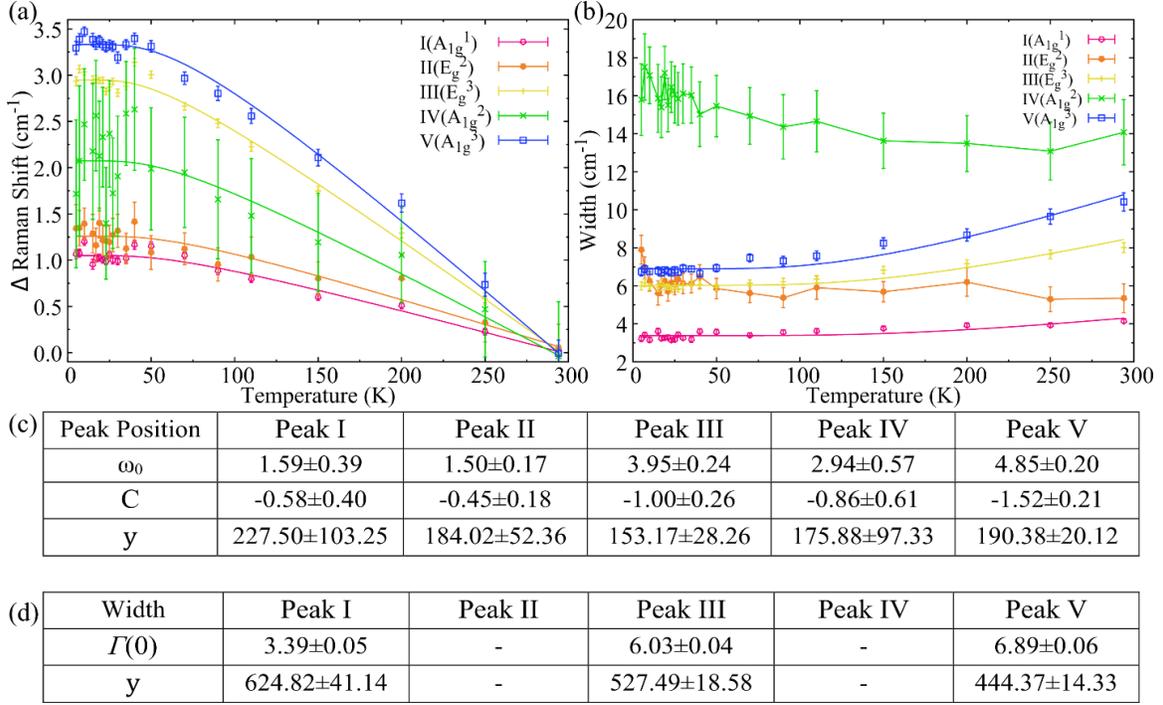

(c)

| Peak Position | Peak I | Peak II | Peak III | Peak IV | Peak V |
|---|---|---|---|---|---|
| $\omega_0$ | 1.59±0.39 | 1.50±0.17 | 3.95±0.24 | 2.94±0.57 | 4.85±0.20 |
| C | -0.58±0.40 | -0.45±0.18 | -1.00±0.26 | -0.86±0.61 | -1.52±0.21 |
| y | 227.50±103.25 | 184.02±52.36 | 153.17±28.26 | 175.88±97.33 | 190.38±20.12 |

(d)

| Width | Peak I | Peak II | Peak III | Peak IV | Peak V |
|---|---|---|---|---|---|
| $\Gamma(0)$ | 3.39±0.05 | - | 6.03±0.04 | - | 6.89±0.06 |
| y | 624.82±41.14 | - | 527.49±18.58 | - | 444.37±14.33 |

**Figure S5.** Temperature dependent Raman peak positions, relative to those measured at room temperature (panel a); and their corresponding widths (panel b), determined from the Lorentzian fits of Figure 3(a) in the main text. Solid lines indicate the anharmonicity model [ref. 35 in the main text] given by the equations below. Panel c and d show the fitting parameters for panel a and b, respectively.

$$\Omega(T) = \omega_0 + C\left(1 + \frac{2}{e^{y/T} - 1}\right), \text{where } y = \frac{\hbar\omega_0}{2k_B}$$

$$\Gamma(T) = \Gamma(0)\left[1 + \frac{2}{e^{y/T} - 1}\right], \text{where } y = \frac{\hbar\omega_0}{2k_B}$$

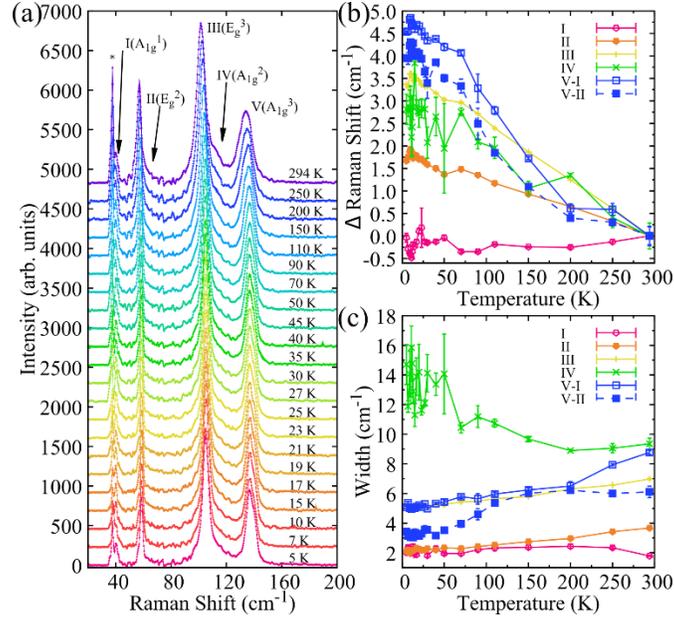

**Figure S6.** Temperature dependences and intralayer spin-phonon coupling of MnBi$_4$Te$_7$. Panel (a) illustrates the cryogenic Raman spectra. The peak marked with * is from the laser line. We used Lorentzian fit to extract the Raman peak positions (panel b) and the widths (panel c). Panel (b) represents the deviation of the Raman peak positions from the ones at room temperature. Peaks V-I and V-II are the Davydov pair that we discuss in the main text. Panel (c) widths of the Raman peaks determined from the Lorentzian fits.

| Mode | MnBi$_2$Te$_4$ | | MnBi$_4$Te$_7$ | | MnBi$_6$Te$_{10}$ | MnBi$_8$Te$_{13}$ |
|---|---|---|---|---|---|---|
| | Calculated | Measured | Calculated | Measured | Measured | Measured |
| $E_g$ | 30.3 | - | 27.5 | 27.0±0.4 | - | - |
| $E_g$ | - | - | 38.4 | - | - | - |
| $A_{1g}$ | 46.0 | 45.7±0.1 | 41.5 | 40.6±0.1 | 40.6±0.1 | 40.1±0.2 |
| $A_{1g}$ | - | - | 58.8 | 57.0±0.1 | 58.6±0.1 | 59.4±0.1 |
| $E_g$ | 74.5 | 66.9±0.3 | 68.9 | 57.8±0.1 | - | - |
| $E_g$ | 109.6 | 103.4±0.1 | 109.3 | 101.7±0.1 | 101.1±0.1 | 101.1±0.1 |
| $A_{1g}$ | 117.4 | 113.5±0.6 | 115.5 | 112.4±0.3 | - | - |
| $A_{1g}$ | - | - | - | 132.1±0.2 | - | - |
| $A_{1g}$ | 144.1 | 138.6±0.1 | 141.3 | 136.3±0.2 | 132.6±0.1 | 132.9±0.1 |

**Table S1.** Calculated and measured peaks from Raman spectra of MnBi$_{2n}$Te$_{3n+1}$ ($n$=1,2,3,4). Around 38 cm$^{-1}$ peak overlaps with the laser-induced peak and therefore, it cannot be measured. All units are in cm$^{-1}$.